\documentclass[]{aa}     
\usepackage{graphicx}
\usepackage{times}
\usepackage{natbib}
\bibpunct{(}{)}{;}{a}{}{,} 
\def\cm2{\,{\rm cm^{-2}}}

\def\13co{\,{\rm ^{13}CO}}
\def\h2{\,{\rm H$_{2}$}}
\def\hi{H\,{\sc i~}} 

\def\arcmin{$^{\prime}$~}
\def\arcsec{$^{\prime\prime}$~}
\def\aap{{\rm A\&A}}

\def\apj{{\rm ApJ}}

\def\apjs{{\rm ApJS}}
\def\apjl{{\rm ApJL}}

\def\mnras{{\rm MNRAS}}

\begin{document}
 
\title{A submillimeter exponential disk in M~51: evidence for an
extended cold dust disk}
 
   \subtitle{}
 
\author{R. Meijerink,
          \inst{1}
        R.P.J. Tilanus,
          \inst{2}
        C.P. Dullemond,
          \inst{3}
        F.P. Israel,
          \inst{1}
        and P.P. van der Werf
          \inst{1}
           }
 
\authorrunning{R. Meijerink et al.}
\titlerunning{A submillimeter exponential disk in M~51}

   \offprints{R. Meijerink}
 
  \institute{Sterrewacht Leiden, P.O. Box 9513, 2300 RA Leiden,
             The Netherlands
  \and       Joint Astronomy Centre, 660 N. A'ohoku Pl., Hilo,
             Hawaii, 96720, USA
  \and       Max-Planck-Institut f\"ur Astrophysik, Karl-Schwarzschild-Str. 1,
             Postfach 13 17, 85741 Garching}

\date{Received ????; accepted ????}
 
\abstract{A 850~$\mu$m map of the interacting spiral galaxy M~51 shows
well-defined spiral arms, closely resembling the structures seen in CO
and HI emission. However, most of the 850~$\mu$m emission originates
in an underlying exponential disk, a component that has not been
observed before in a face-on galaxy at these wavelengths. The
scale-length of this disk is 5.45 kpc, which is somewhat larger than
the scale-length of the stellar disk, but somewhat smaller than that
of atomic hydrogen. Its profile can not be explained solely by a
radial disk temperature gradient but requires the underlying dust to
have an exponential distribution as well. This reinforces the view
that the submm emission from spiral galaxy disks traces total hydrogen
column density, i.e.\ the sum of \h2 and \hi. A canonical gas-to-dust
ratio of 100$\pm$26 is obtained for $\kappa_{850}=1.2\ \rm g^{-1} \rm
cm^{2}$, where $\kappa_{850}$ is the dust opacity at 850~$\mu$m.}

\maketitle
 
\section{Introduction}

After the introduction of the SCUBA bolometer camera at the JCMT a
number of nearby galaxies have been imaged in the submm range to
investigate the cold dust component of the ISM \citep{Alton2002,
Israel1999, Stevens2000}. The spatial distribution of dust and its
relation to other components of the ISM is best studied in face-on
systems, but at submm wavelengths surface brightnesses are typically
low. This particularly hampers the search for diffuse emission from
inter-arm regions. Moreover, most of these nearby galaxies are larger
than the field-of-view of SCUBA requiring scan-mapping which, as
explained below, causes features in the background at the level of a
few milli-Jansky frustrating further attempts to determine the level
of any weak extended emission. The example of NGC~6946 is
illustrative. Little diffuse emission in the inter-arm region was seen
by \citet{Alton2002} in the 850~$\mu$m map. Consequently, they state
that "relatively little is known about how the dust is distributed
with respect to the spiral arms and, in particular, whether inter-arm
grain material is prevalent or not".

At resolutions much lower than those provided by the JCMT, the cold
dust in galaxies has been mapped with ISO in the far-infrared
\citep{Haas1998, Popescu2002, Hippelein2003, Tuffs2003,
Popescu2003}. These observations confirm the ubiquitous presence of
cold dust below 20~K first observed by \citet{Chini1986} and indicate
that the cold dust component is smoothly distributed over the disk and
heated by the diffuse interstellar radiation
\citep{Hippelein2003,Xu1994}.  \citet{Alton2001} reported evidence for
diffuse inter-arm dust in SCUBA observations of NGC~7331. For a recent
discussion of `dusty disks' see also \citet{Bianchi2004}.

M~51 is a well studied, face-on interacting spiral galaxy at an
assumed distance of 9.7 Mpc. It is well known for its strong CO
emission, but contrary to NGC~6946 its density-wave and spiral arms
are well-organized and prominent, showing a high contrast in the radio
continuum (see e.g. Tilanus \& Allen
\citeyear{Tilanus1991,Tilanus1989}). The distribution of cold dust in
the nuclear region of M~51 was previously mapped at 1.2 mm by
\citet{Guelin1995}, who noted a close correlation with CO line
emission. Our current 850~$\mu$m observations (Fig.~\ref{M51pict})
show that its dust emission is exceptionally strong. The spiral arms
are clearly distinguished, but there is also a substantial amount of
emission from an extended diffuse disk. To our knowledge, such a
submillimeter disk has not been clearly identified in any other
galaxy, but based on the ISO observations discussed previously, it may
be a common component of spiral galaxies. In this paper, we will
explore the nature and properties of this disk. The 850~$\mu$m spiral
structure of M~51 will be discussed in a future paper.

\section{Observations and data reduction}

M~51 was observed at 850~$\mu$m and 450~$\mu$m in the spring of 1998
and 1999 at the JCMT\footnote{The James Clerk Maxwell Telescope is
operated by the Joint Astronomy Centre on behalf of the Particle
Physics and Astronomy Research Council of the United Kingdom, the
Netherlands Organization for Scientific Research and the National
Research Council of Canada} using the Submillimeter Common User
Bolometer Array (SCUBA: \citet{Holland1999}). In order to map a
13.5\arcmin x 13.5\arcmin region around M~51, the camera was used in
scan-map mode during which it is scanned at a rate of 3\arcsec per
second across the field, while the secondary was chopped with a
frequency of 7.8 Hz in right ascension or in declination to cancel
atmospheric signal variations as well as bolometer DC
fluctuations. This uses the revised version of the
Emerson-Klein-Haslam algorithm proposed by Emerson
(\citeyear{Emerson1995}).  The scan-angle across the field,
(15.5$\pm$60$^{\rm o}$), was chosen such that the resulting image is
fully sampled. The total integration time was about 20 hours spread
over 6 nights.

To restore the brightness distribution of the source, the chop must be
deconvolved from the observed map using a division in the Fourier
plane \citep{Jenness2000}. The Fourier-transform (FT) of the chop is a
sine wave with zeroes at the origin and at harmonics of the inverse
chop throw. The initial observations used chop-throws of 20, 30, and
65 arcsecs. To minimize regions of low weight (close to zeroes) in the
Fourier plane we later added throws of 44 and 68 arcsecs. Each
individual observation was flat-fielded, corrected for atmospheric
opacity, and despiked. For each of the 10 different
chop-configurations the individual maps were coadded. \ Next,
the resulting 10 maps were Fourier transformed, weighted, and coadded.
The coadded image was transformed back to yield the finished
image. This is the standard reduction for SCUBA scan-map observations
as implemented in the SCUBA User Reduction Facility software (SURF:
Jenness \& Lightfoot \citeyear{Jenness1998}).

\begin{figure}[ht]
\centerline{\includegraphics[height=80mm,clip=]{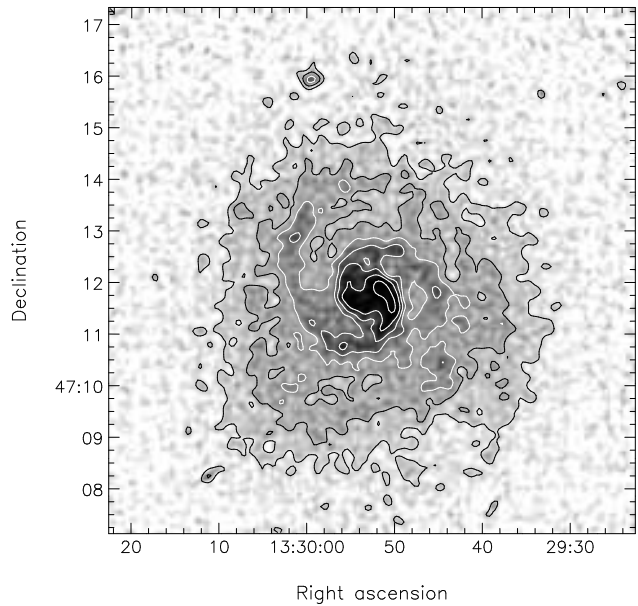}}
\caption{SCUBA calibrated and background corrected 850~$\mu$m
grey-scale image of M~51 with 850~$\mu$m contour superimposed. The
contour levels are 13.5 (=1.6 $\sigma$), 27, 40.5, 58.5, 72, 90 and
117 mJy/beam. 13.5 mJy/beam corresponds to 2.6 MJy~sr$^{-1}$. This
image clearly shows the extended 850~$\mu$m disk. The small source to
the north is the nucleus of NGC~5195.}
\label{M51pict}
\end{figure}

The data were flux-calibrated using Mars and the secondary standards
HL~Tau and CRL~618. In the final image, the beam-size at 850~$\mu$m is
about 15\arcsec (FWHM).  The calibrated image is shown in
Fig.~\ref{M51pict} and has an overall rms of
$\sim$9~mJy/beam. The inconspicuous source to the north is the
nucleus of NGC~5195.  The familiar spiral pattern is clearly visible
and can be traced over a large fraction of the disk. This is quite
unlike the submm continuum observations of other face-on spirals,
which in general lack the sensitivity to trace the arms outside the
inner region. In detail, the morphology of the arms resembles closely
the one seen in CO emission \citep{Aalto1999,Wielebinski1999}, and we
will address the contamination of the continuum emission by in-band
line emission in section 3.

Remarkably, the 850~$\mu$m emission from M~51, showing the
distribution of the cool dust, is dominated by a diffuse exponential
disk, a component that has not been directly imaged before at sub-mm
wavelengths and with a high resolution in a face-on spiral. The
450~$\mu$m observations also show spiral structure and an extended
disk. In part because of the difficulty to derive an accurate flux
calibration and the poorer image quality, we will not further discuss
the 450~$\mu$m data in this paper.

\subsection{Sky noise and background removal}

Analyzing extended structures in SCUBA scan maps is a delicate process. 
While there is no doubt about the reality of the exponential disk, 
careful processing is required for realiable parameter extraction. There 
are a number of reasons for this.

Firstly, while regions of low weight have been minimized through the
use of many different chop configurations, the zero near the origin of
the Fourier plane can not be suppressed. This results in a poor
determination of the largest-scale features in the map, including the
total flux-density. The effect is similar to that caused by missing
short-spacings in radio interferometry observations and it has a
similar effect on the background.

\begin{figure}[]
\centerline{\includegraphics[height=80mm,clip=]{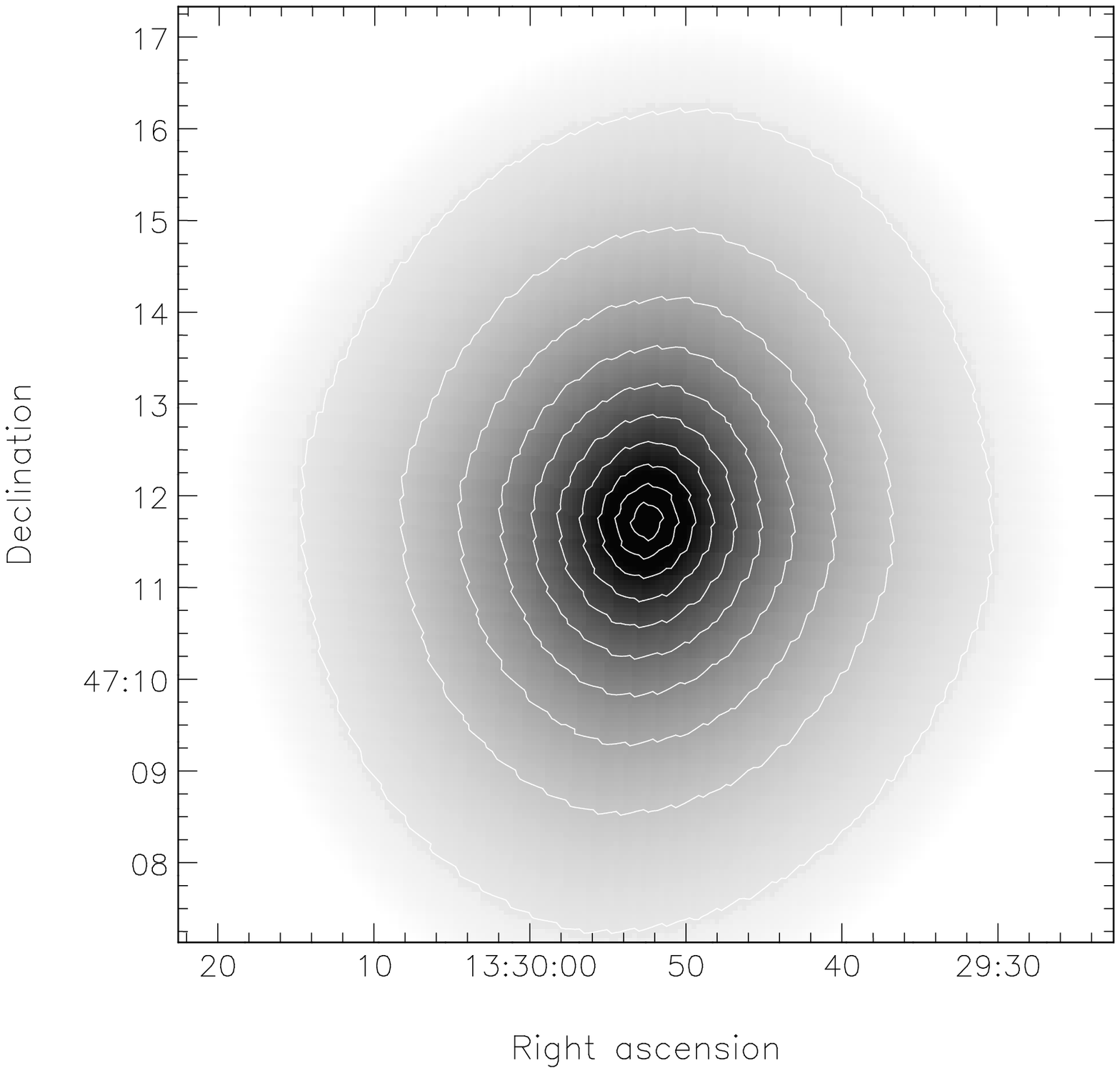}}
\centerline{\includegraphics[height=80mm,clip=]{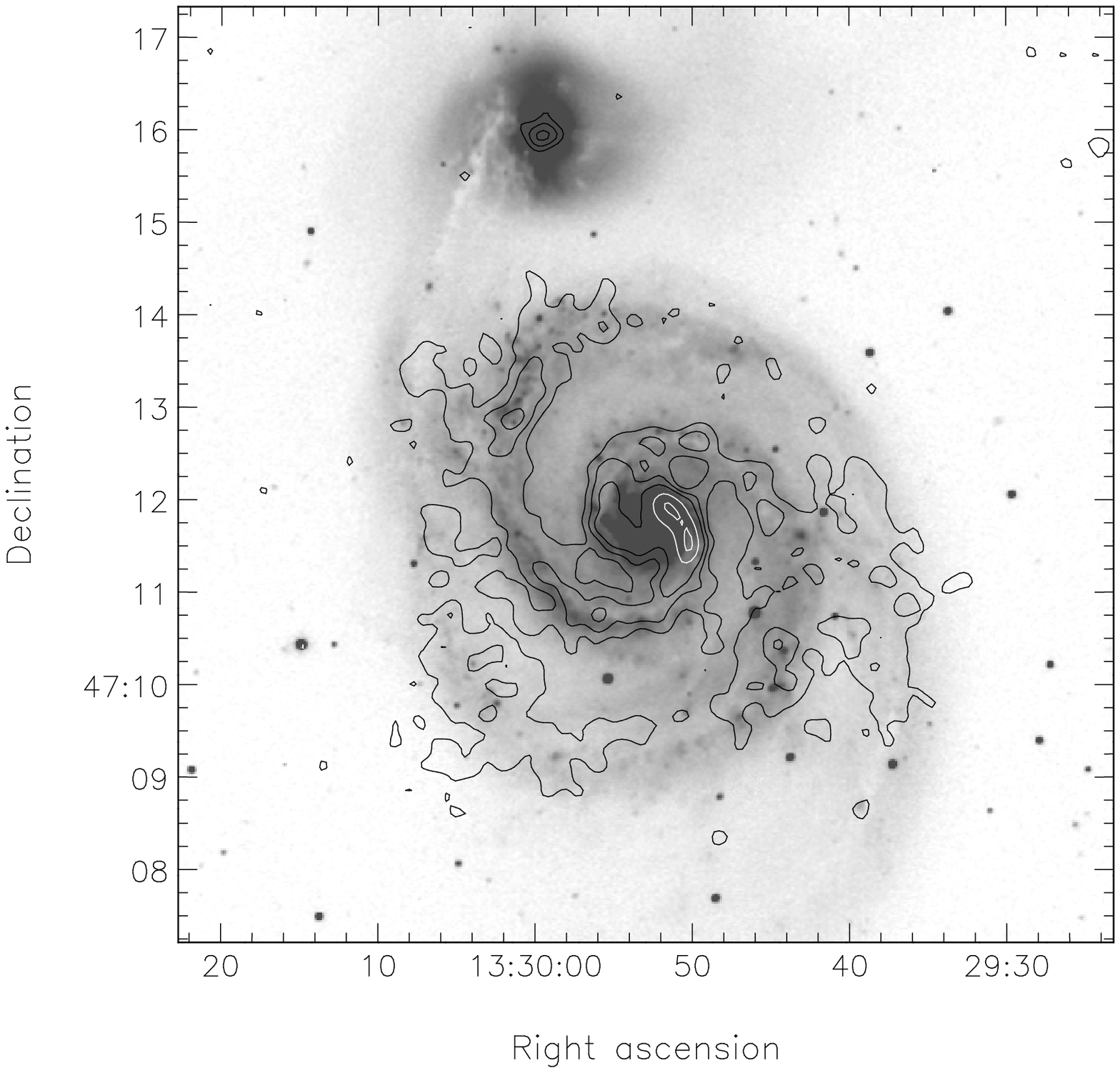}}
\caption{Top: Fitted exponential disk subtracted from the total
850~$\mu$m emission as shown in Fig. \ref{M51pict}. Contour steps are
6~mJy/beam (1.15~MJy sr$^{-1}$). The maximum intensity of the disk is
66~mJy/beam with a major axis scale-length of
1.95\arcmin. Bottom: Contours of 850~$\mu$m in M~51 after
subtraction of the exponential disk superimposed on a 6400\AA\ optical
image \citep{Tilanus1988}. Contours levels are 12 (=1.5 $\sigma$;
2.3~MJy~sr$^{-1}$), 24, 36, 56 and 72 mJy/beam. The image shows the
centroid of the 850~$\mu$m emission to be located along the inner
edges of the spiral arms and the optical dust lanes. }
\label{exponent}
\end{figure}

\begin{table} 
\caption[]{Galaxy parameters} 
\begin{flushleft} 
\begin{tabular}{lll} 
\hline 
\noalign{\smallskip} 
			    & M~51				  & \\
\noalign{\smallskip}
\hline
\noalign{\smallskip}
Type$^{a}$     	            & SAS(s)bcp				  & \\
Radio Centre:		    & 					  & \\
R.A. (B1950)$^{b}$ 	    & 13$^{h}$27$^{m}$46.3$^{s}$  	  & \\
Decl.(B1950)$^{b}$          & +47$^{\circ}$27$'$10$''$ 	          & \\
R.A. (J2000)$^{b}$          & 13$^{h}$29$^{m}$52.7$^{s}$  	  & \\
Decl.(J2000)$^{b}$          & +47$^{\circ}$11$'$42$''$ 	          & \\
$V_{\rm LSR}^{c}$           & +464~km~s$^{-1}$                    & \\
Inclination $i^{c}$ 	    & 20$^{\circ}$ 			  & \\
Position angle $P^{c}$      & 170$^{\circ}$ 			  & \\
Distance $D^{d}$            & 9.7 Mpc 				  & \\
Scale           	    & 21$''$/kpc 			  & \\
Scale-lengths (disk):       & (arcsec)                            & (kpc) \\
B$^e$                       & 92.8$\pm$9.4                        & 4.36$\pm$0.45\\  
I$^e$                       & 81.2$\pm$7.0                        & 3.81$\pm$0.33\\
R$^e$                       & 80.4$\pm$6.1                        & 3.77$\pm$0.29\\
K$^e$                       & 87.1$\pm$6.8                        & 4.09$\pm$0.32\\
Effective radius (bulge):   & (kpc)                               & \\
B$^f$                       & 0.94$\pm$0.72                       & \\
V$^f$                       & 0.98$\pm$0.52                       & \\
R$^f$                       & 1.07$\pm$0.57                       & \\
I$^f$                       & 1.06$\pm$0.50                       & \\
\noalign{\smallskip}
\hline
\end{tabular}
\end{flushleft}
Notes to Table 1:\\
$^{a}$ RSA \citep{Sandage1987} \\
$^{b}$ \citet{Turner1994} \\
$^{c}$ \citet{Tully1974} \\
$^{d}$ \citet{Sandage1975} \\
$^{e}$ \citet{Beckman1996} \\
$^f$ \citet{Laurikainen2001} \\
\label{galaxyparam}
\end{table}

Secondly, chopping removes noise due to sky fluctuations with a
frequency comparable to the chop-frequency. It cannot remove the
effect of a sky signal which is steadily increasing or decreasing
while scanning from one side of the field to the other, i.e. a slowly
and monotonically varying component of the sky. In such situations the
`on' minus `off' will always leave a
small residual sky signal. We have used the standard SURF option, {\it
scan\_rlb}, to subtract a linear baseline from each bolometer-scan
fitted to the extreme 2 arcmin on each side of the scan.  While this
significantly improves the flatness of the background, it will leave
any ripples that may be due to {\it non-linear} variations of the sky
during each scan. To further improve the sky-noise removal we have
used the SURF routine {\it calcsky} \citep{Jenness1998}, which is
similar to the {\it remsky} routine used for non-raster data. A
detailed discussion of these methods is beyond the scope of the paper,
but while {\it scan\_rlb} relies on the fact that there is no source
signal at the edges of the map, {\it calcsky} makes use of the fact
that (a component of) the sky noise is correlated across the bolometer
array during each integration step.  The application of both methods
improved the flatness of the background without changing the overall
appearance of the image.

Nevertheless, even with the application of the above, relatively
objective, methods background artifacts remained, likely because of
the missing Fourier zero. This is a common feature of SCUBA scan maps.
We used an iterative procedure to correct the image which involved
subtracting an inclined exponential (the diffuse disk), un-sharp
masking and blanking of the remaining source signal (arms, nucleus),
followed by a polynomial plane fit and convolution to obtain a smooth,
large-scale representation of the background. The operations were done
using the finished coadded image. While comparable to the procedure
used by e.g.\ \citet{Alton2002}, it is subjective in nature. As
guidelines we used the following criteria: the flatness of the
background and the assumption that the inter-arm emission originates
from the diffuse disk only.

The exponential diffuse disk will be discussed separately in section
4: we fitted a scale-length of 1.95\arcmin, perhaps fortuitously the
same as found for the 20-cm `base' disk \citep{Tilanus1988}, the
inclination and position angle as in Table~\ref{galaxyparam}, and
fitted a peak of 66~mJy/beam.  Across the disk of the galaxy the
fitted background approximates a linear ramp of $-12\pm10$~mJy/beam:
from $-2$ in the south to $-22$ in the north.  The mean level of
$-12$~mJy/beam corresponds to (minus) $9\%$ of the 132 mJy/beam peak
in the raw map. By inspecting the resulting inter-arm regions
(Fig.~\ref{exponent}) we estimate the local background to be accurate
to a level of about $\pm$2~mJy/beam or about $\pm$5\% of the typical
knots in the arms. Given the overall rms in the image of 9~mJy/beam,
this uncertainty is about four times larger than the statistical
uncertainty of the mean over the areas inspected. Fig.~\ref{M51pict}
shows the calibrated observations after correcting for the low-level
large-scale background fluctuations.

\subsection{Total flux density}

Using an aperture of 8\arcmin, which just excludes the nucleus of
NGC5195, we find an integrated flux of $14.5 \pm 1.1$~Jy. The
uncertainty quoted indicates the range of values found when varying
the aperture from 7.4\arcmin to 10\arcmin, not the uncertainty
resulting from the background variations. To estimate the latter we
inspected the variation of the mean level in 2.5\arcmin regions around
M~51, similar to the size of the nucleus and inner arms. This
variation is about 0.5~mJy/beam and can serve as an indication of the
quality of the background correction where it applies to regions
already largely free from emission. By contrast we note that the
$\pm$2~mJy/beam uncertainty of the background in the inter-arm
regions, as mentioned in the previous section, is positive in the
south and negative in the north, suggesting that the uncertainty of
the zerolevel taken over the whole of the disk of M~51 is less than
2~mJy/beam. Based on these considerations we adopt an overall
uncertainty in the background level of 1.5~mJy/beam corresponding to
1~Jy when summed over the 8\arcmin aperture. Finally we need to add a
10\% uncertainty in the flux of the calibrators at submm
wavelengths. Treating the various uncertainties as statistical errors,
we derive a total 850~$\mu$m flux for M~51 of $15 \pm 2$~Jy. This
value includes the contribution from line emission within the passband
of SCUBA's 850~$\mu$m filter, a topic which will be discussed in the
next section.

\section{Contamination by J=3-2 CO line emission}

The broad passband of the SCUBA instrument includes the wavelength of
the $J$=3-2 $^{12}$CO line. Thus, at any point in the map the
continuum emission measured by SCUBA is, in principle, contaminated by
CO line emission.  In order to investigate the extent of the
contamination we obtained with the JCMT in spring 2001 a set of
CO(3-2) spectra across both inner arms of M~51 and supplemented these
with archived spectra of the nuclear region. In the arms, the CO line
contribution can be up to 30-40 percent of the observed continuum
emission {\it after subtraction of the exponential disk}. Thus, a
substantial fraction of the arm morphology seen in the SCUBA map
directly results from contaminating line emission. Because of the
apparently close correspondence between CO and dust emission
\citep{Guelin1995}, the effect is mostly quantitative and does not
change the overall morphology of the arms qualitatively.  The $J$=3--2
emission from the inter-arm regions is weak and in individual
pointings does not exceed the noise in the measurements. In the center
of M~51, we find an upper limit of 12 per cent ($3\sigma$) to any line
contribution to the {\it emission from the fitted disk}. Consequently,
we may assume either a constant CO contribution at any level below 12
per cent, or a maximum contribution of 12 per cent at the center
decreasing to zero at the outer edge. These two possibilities thus
define lower and upper limits, respectively, to the 850$\mu$m emission
scale-length.

\section{The diffuse disk}

We have separated emission in the diffuse disk from arm emission by
subtracting an inclined exponential disk and making the assumption
that the inter-arm emission originates from the diffuse disk only (see
also section 2.1). The best fit to the data is given by an exponential
disk with a maximum intensity of 66 mJy/beam, and a major axis
intensity scale-length of 1.95\arcmin, corresponding to a linear
scale-length of 5.45~kpc at the assumed distance of M~51. As
1~MJy~sr$^{-1}$ = 5.22~mJy/beam, the maximum intensity 66 mJy/beam =
12.7~MJy~sr$^{-1}$. The subtracted exponential disk and the arm
emission are shown in Fig. \ref{exponent}. The global parameters
adopted for M~51, such as the distance, inclination, and position
angle, are given in Table~\ref{galaxyparam}.

The apparent existence of a diffuse exponential disk raises further
interesting questions. What kind of dust distribution can account for
such a disk: an exponential dust distribution at a constant
temperature, a constant dust distribution with a temperature gradient,
or both an exponential distribution {\it and} a temperature gradient?
What is the dust mass in proportion to the amount of gas?

\subsection{The Monte Carlo radiation transfer code RADMC}

To answer these questions, we have modeled the observed emission using
a modified version of the axi-symmetric Monte Carlo code {\tt RADMC}
(see Dullemond \& Dominik \citeyear{Dullemond2004}). This continuum
radiative transfer code is based on the method of \citet{BenW} and
\citet{Lucy1999}, but adapted to suit the purpose of this paper. The
basic idea of this method is as follows. The input luminosity of the
system comes from the population of stars in the galaxy. At every
location in the galaxy this is represented by a wavelength-dependent
source function, with the spectral shape of the local stellar
population. Integrated over wavelength and over space, this source
function yields the total intrinsic bolometric luminosity of the
galaxy.  Each spatial cell represents a certain fraction of this
luminosity, dependent on the spatial distribution of the stars. And
within such a cell each wavelength interval represents again a
fraction of this, dependent on the local spectral shape of the stellar
population. In this way each space-wavelength cell represents a
well-defined fraction of the total input energy into the problem.

The total input luminosity is now evenly divided over $N_{\mathrm{phot}}$
photon packages. These packages are launched into the galaxy at random
angles from randomly chosen space-wavelength cells, where the probability
for each space-wavelength cell to launch this photon is proportional to
the fraction of the total input luminosity represented by that cell.
Each photon package is launched and performs a random walk through the
computational domain until it eventually escapes to infinity. After that,
the next photon package is launched.

As these photon packages travel through the computational domain they
can experience scattering and absorption events. A scattering event
merely changes the direction of the photon package, but an
absorption-reemission event can also change its wavelength. The
probability function for this random wavelength is
$dP(\lambda)/d\lambda \sim \partial B_\lambda(T)/\partial T$, where
$T$ is the local dust temperature at the moment the photon package
enters the cell. The dust temperature changes during the simulation,
increasing with the number of photon packages that have entered the
cell. Each photon package travelling through a cell increases the
`energy' of the cell proportional to the energy of the photon package,
the length $l$ of the path through the cell and the opacity of the
cell at that wavelength.  The temperature of the cell is increased
after each passage in such a way that the total dust emission of the
cell $4 \pi \int_0^\infty \rho_{\mathrm{dust}}\kappa_\lambda
B_\lambda(T)d\lambda\times V_{\mathrm{cell}}$ (where
$V_{\mathrm{cell}}$ is the cell volume) equals the sum of all
contributions $\rho_{\mathrm{dust}}\kappa_\lambda l
L_{tot} / N_{phot}$ of all photons that have so far passed through
the cell.

The increase of the energy of the cell happens each time a photon travels
through the cell, while a discrete scattering or absorption-reemission event
(which changes the direction and/or wavelength of a photon package) occurs
only at random locations along the path of the photon package. This random
location is chosen to be at an optical depth of
$\tau=|\log(\mathrm{ran}(i_{\mathrm{seed}}))|$ away from the last discrete
event, where $\mathrm{ran}(i_{\mathrm{seed}})$ is a random number between 0
and 1.

Once all the photons have been launched and have left the system, the dust
temperature and scattering source function at every location in the galaxy
has been determined. Modulo random fluctuations this should be the correct
temperature and scattering source term distribution. Now we use a
ray-tracing code (part of the radiative transfer package {\tt RADICAL}
\citep{Dullemond2000}) to produce the desired images and/or spectra.
But to verify whether this temperature and scattering source term
distribution is indeed the correct solution, {\tt RADMC} has been tested in
two ways. First of all, it has been verified that flux is conserved to a
high level of precision ($<1$\%) by making spectra at many inclination
angles and integrating over wavelength and inclination to obtain the total
output luminosity. Since the spectra are produced using a ray-tracing code,
and not by counting the outcoming photon packages (as is the case for the
Bjorkman \& Wood code), this energy conservation is not numerically
guaranteed, and therefore represents a good check on the
self-consistency. The code has also been tested against many other codes in
a 2-D radiative transfer comparison project \citep{Pascucci2004}.

\subsection{Assumed disk structure}

We assumed that the stellar distribution consists of two components,
the exponential disk and the bulge. For the bulge we used a Hubble
profile \citep{Ren,Hubble}, which models the stellar emissivity as:

\begin{eqnarray} 
{\cal L}(R,z) & = & {\cal L}_{\rm disk}\,\, \exp\left(-\frac{R}{h_s}-\frac{|z|}{z_s}\right)
       \nonumber\\ & + & \,\, {\cal L}_{\rm bulge}(1+B^2)^{-3/2}
\end{eqnarray} 
  
\noindent
in which $B$ is given by the expression:

\begin{eqnarray}
B=\frac{[R^2 + (z/(b/a))^2]^{-3/2}}{R_e}
\end{eqnarray}

\noindent
where $R$ en $z$ are cylindrical coordinates, ${\cal L}_{\rm bulge}$
and ${\cal L}_{\rm disk}$ bulge and disk emissivities at the center of
the galaxy respectively, $h_s$ and $z_s$ the respective stellar
scale-lengths and scale-heights, $R_e$ the effective bulge radius and
$b/a$ the minor/major axial ratio.

We explored both exponential and constant dust distributions. The
parameterization to describe the exponential dust distribution is the
same as used for NGC~891 by \citet{Xilouris}:

\begin{equation}
\rho(R,z)=\rho_c \,\,\exp\left(-\frac{R}{h_d}-\frac{|z|}{z_d}\right)
\end{equation}

\noindent
with $\rho_c$ being the central dust density, $h_d$ the dust
distribution scale-length and $z_d$ the dust distribution
scale-height.  A constant dust distribution was obtained from the same
expression by putting the scale-length at infinity.

Extinction coefficients were obtained from the model described by
\citet{DraineLee}. We considered four different silicate-to-graphite
ratios in the grains, 0.5:0.5, 0.4:0.6, 0.3:0.7 and 0.2:0.8. The
particle sizes $a$ were set between 0.005 and 0.25~$\mu$m with a size
distribution proportional to $a^{-3.5}$. Our modeling did not include
either PAH's or very small dust particles. Although these particles
dominate the mid-infrared emission at e.g. 10--20~$\mu$m, their
contribution to submillimeter emission at 850~$\mu$m is expected to be
negligible (see e.g. Li \& Draine \citeyear{Li2001}).

The spectral energy distribution of the stars was taken from the
models calculated by \citet{BruzualCharlot}. Specifically, we used the
model for a continuously star-forming galaxy with a star-forming rate
of $10^{-10}$ M$_\odot/\rm{yr}$ per unit solar mass, assuming a galaxy
age of 10 Gyr. We verified that decreases in the assumed age to 5 Gyr
or even 1 Gyr do not cause significant changes in the parameters
needed to fit the 850~$\mu$m radial profile. The dust temperature
rises marginally, causing a small change in the dust surface
density. The latter, however, is much more influenced by the
uncertainty in the total luminosity.

\section{Results and analysis}

\subsection{M~51 stellar parameters}

First, the scale-length of the stellar emission must be determined as
accurately as possible, because it dominates determinations of the
scale-length of the 850~$\mu$m dust emission. The stellar emissivity
of M~51 has been extensively studied by \citet{Beckman1996} who
determined both arm and inter-arm scale-lengths in the $B$, $R$, $I$
and $K$ photometric bands, where they did not correct for
extinction. Their inter-arm results are summarized in
Table\,\ref{galaxyparam}. Emission in the $R$ and $I$ bands
predominantly traces old stars which are distributed more or less
homogeneously through the disk. We used these bands to determine the
scale-length of the stellar emission. As this emission is susceptible
to dust attenuation, we have taken extinction into account as
well. Starting with an initial estimate of the intrinsic scale-length
of the stars, the radiation transfer code computes a value for the
attenuated scale-length of the stars, which was then compared with the
observations.

The M~51 bulge effective radius was not determined. However, as it
does not dominate the radial profile, we used a value of 1.03~kpc
characteristic for similar galaxies \citep{Laurikainen2001}, also
summarized in Table\,\ref{galaxyparam}.

Unlike the scale-length of the exponential star distribution, the
scale-heights of both the stellar disk and the dust are in effect
degenerate: model and data do not provide useful constraints.  For
instance, a wide range of stellar scale-heights yields almost
identical 850~$\mu$m emission profiles. Dust temperatures are
marginally affected, requiring some changes in the 10 -- 100~$\mu$m
wavelength range of the spectrum. We kept the ratio $h_s/z_s$
identical to that found in NGC~891 \citep{Xilouris}. Obviously, as
long as the face-on optical depth is kept constant the submillimeter
emission does not change perceptibly either when we vary the dust
scale-height. When we increase dust scale-heights from 0.25 to 0.35
kpc, temperatures change by no more than 1-2 degrees.

In order to determine the total luminosity of M~51, we integrated the
UV (OAO), visual and far-infrared (IRAS) flux-densities as provided in
the NASA/IPAC Extragalactic Database, the 170 $\mu$m point as given by
\citet{Tuffs2003} and our 850 $\mu$m point. We find a luminosity of
$L_{tot} = 1.1\times 10^{11} L_\odot$, which corresponds to a
starformation rate of a few M$_\odot$yr$^{-1}$, with an estimated
uncertainty of about $30\%$. Of this total luminosity, about $5\%$
originates in the bulge and the remaining $95\%$ in the exponential
disk.

\subsection{Disk parameter determination}

Even when the stellar luminosity and the CO contamination would be
known, the stellar scale-height, $z_s$, and the dust scale-height,
$z_d$, are degenerate. Adopting values as outlined in the previous
section, we explored the parameter space over three variables: (i) the
total amount of dust, which scales with the central face-on optical
depth $\tau_{V,f}$ integrated from infinity to the galaxy center, (ii)
the scalelength of the dust distribution $h_d$, and (iii) the
scale-length of the stellar disk $h_s$, since the observed
scale-length is attenuated by dust. These variables are uniquely
determined when the stellar luminosity and the CO contamination are
known. Since this is not the case, we identified combinations of
parameters best reproducing the observed 850~$\mu$m radial profile.

\begin{figure}[]
\centerline{\includegraphics[height=65mm,clip=]{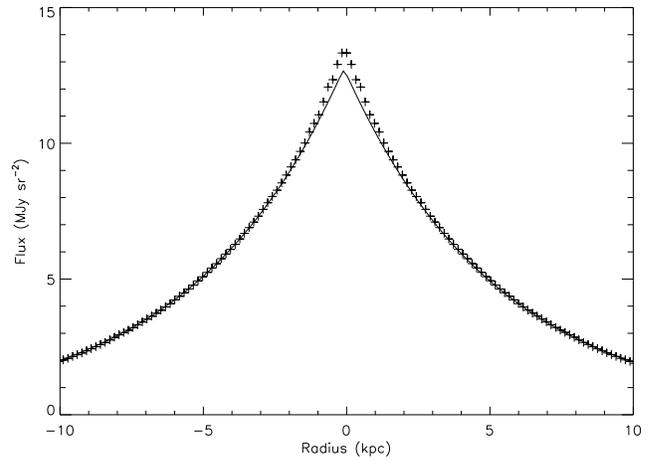}}
\caption{The model exponential disk (crosses) compared to the actual
emission of the M~51 disk (solid line), for $h_d=7.65$~kpc, $h_s=3.15$~kpc,
$L=1.1\cdot 10^{11}\ L_\odot$, $\tau_{V,f}=4.35$ and no CO
contamination.}
\label{exponentialfit}
\end{figure}

\begin{figure}[]
\centerline{\includegraphics[height=65mm,clip=]{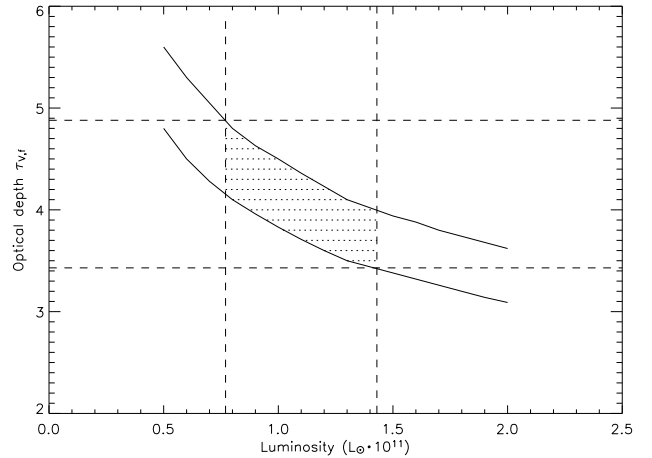}}
\caption{Resulting optical depth for different luminosities assuming
either zero CO contamination (upper line) or an omnipresent maximum 12
per cent contamination (lower line). The shaded area shows the
possible solutions.}
\label{tauplot}
\end{figure}

We summarized the model results in table \ref{modelresults}. An
example fit is shown in Fig.~{\ref{exponentialfit}. Note that the
model provides a poor fit in the very center. This is caused by the
presence of the bulge in the model which provides additional dust
heating, hence more submillimeter emission in the center. We only
subtracted the disk component from the observed emission.

\begin{table}
\caption[]{Model results} 
\begin{flushleft} 
\begin{tabular}{llll} 
\hline
\noalign{\smallskip}
CO contamination                                  & No        & Maximal   & Variable \\
\noalign{\smallskip}
\hline
\noalign{\smallskip}
$h_s$~(kpc)                                       & 3.15      & 3.15      & 3.15   \\ 
$h_d$~(kpc)                                       & 7.65      & 7.65      & 8.65   \\
$\tau_{V,f}$ ($L=0.8\cdot 10^{11}\ L_\odot$)      & 4.8       & 4.1       & 4.2    \\
$\tau_{V,f}$ ($L=1.1\cdot 10^{11}\ L_\odot$)      & 4.4       & 3.7       & 3.7    \\
$\tau_{V,f}$ ($L=1.4\cdot 10^{11}\ L_\odot$)      & 4.0       & 3.4       & 3.5    \\
\noalign{\smallskip}
\hline
\end{tabular}
\end{flushleft}
\label{modelresults}
\end{table}

In Figure \ref{tauplot} we show the required face-on central optical
depth for a range of total galaxy luminosities and CO
contributions. For a luminosity $L_{tot} = 1.1\times 10^{11} L_\odot$
these optical depths $\tau_{V,f}$ range from 3.7 (assuming $12\%$ CO
contamination) to 4.4 (zero contamination). Inclusion of the
luminosity uncertainty increases this to a slightly larger range
$\tau_{V,f}$ 3.4--4.8. The high optical depth derived here is
consistent with the typically higher values derived from far-infrared
and submm compared to optical observations (see e.g. Bianchi
\citeyear{Bianchi2004}). \citet{Popescu2000} found a value of 3.1 for
NGC~891. In this case, however, arm and interarm emission are fitted
simultaneously. The associated scale-lengths vary likewise. For
instance, as long as the CO contamination is constant with radius, we
find a dust scale-length $h_{d}$ = 7.65 kpc and a stellar scale-length
$h_{s}$ = 3.15 kpc. On the other hand, if the CO contaminations is
maximal ($12\%$) at the center and minimal (zero) at a radius of 10
kpc, we obtain a slightly greater scale-length $h_{d}$ = 8.65 kpc. The
scale-length $h_{s}$ remains the same within the uncertainties.The
value of $h_d$ is somewhat smaller than the scale-length of atomic
hydrogen. The value of $h_d$ is comparable to a scale-length of $9\pm
1$ kpc we derive for the atomic hydrogen outside 6~kpc
\citep{Tilanus1991}.

In addition to a radial density gradient, we also find a radial
temperature gradient because the stellar scale-length is much less
than the dust scale-length, i.e. the stars are more centrally
concentrated than the dust. Radial temperature profiles, for the
average grain size at position $z=0$~kpc, are shown in
Fig.~\ref{tempprof}. When the dust scale-length is larger, we also see
a steeper temperature gradient through the galaxy. The small
fluctuations in the Fig.~\ref{tempprof} profile are artifacts due to
statistical noise.

\begin{figure}[]
\centerline{\includegraphics[height=65mm,clip=]{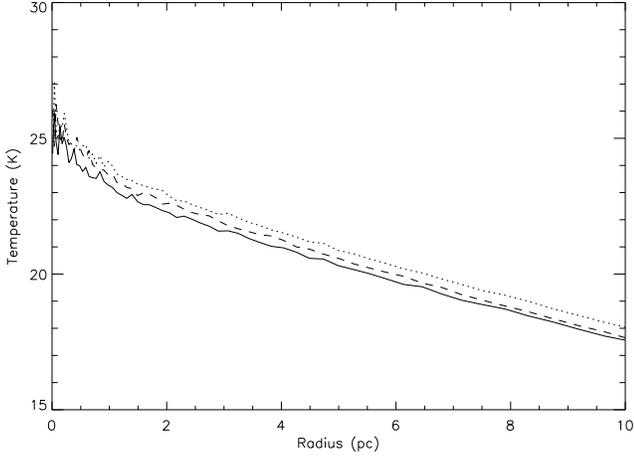}}
\caption{Radial temperature profile for an M~51 total luminosity 
$\rm L=1.1 \cdot 10^{11}\ \rm L_\odot$. The solid line marks zero
CO contamination, the dotted line a maximum 12 per cent contamination,
and the dashed line a contamination decreasing from maximum in the
center to zero at the M~51 outer edge.}
\label{tempprof}
\end{figure}

The good fit of the 850~$\mu$m profile provided by exponential
distributions with finite scale-lengths already suggests that the M~51
disk is hard to explain in terms of a radially constant density
distribution. We nevertheless attempted to force such a fit, and found
that this could only be done if the M~51 stellar scale-lengths were
more than an order of magnitude less than actually found. Thus, a
constant density distribution may be ruled out with confidence. An
exponential density distribution of the cold dust is required in order
to explain to observed exponential 850~$\mu$m disk. 

SCUBA submm observations of the edge-on spiral NGC~891 suggest that
the cold dust distribution traces the total hydrogen column density
i.e. molecular hydrogen in the central regions and neutral hydrogen at
larger radii (Fig. 5 in Alton et al. \citeyear{Alton2000}).
Subsequent far-infrared observations by \citet{Popescu2003} have shown
that the cold dust emission traced the neutral hydrogen into the
extended HI disk. The current observations of M~51, showing an
exponential dust density distribution, a close association with the
molecular hydrogen in the central region, and comparable scale-lengths
of the dust and HI distributions at larger radii further reinforce the
suggestion that the cold dust is correlated with the total hydrogen
column density in galaxies.

\subsection{Dust mass and gas-to-dust ratio}

\noindent Once the density and temperature profiles are determined we
may, in principle, determine the radial dust mass distribution as
well. In Fig.~\ref{surfdens} we show the surface mass densities, as a
function of radius for a graphite and silicate ratio of 0.5:0.5. The
peak dust surface densities are 0.90 (no CO contribution) and 0.77
$\rm M_\odot / pc^2$ (12 percent CO). The central gas (HI + H$_2$ +
He) surface density is $\Sigma_{gas}=60\ \rm M_\odot \rm pc^{-2}$
(calculated in a future paper), with an estimated uncertainty of
25-30\%. This results in a gas-to-dust ratio of 73$\pm$17.

In this dust mixture, we have a dust emission coefficient of
$\kappa_{850}=0.9\ \rm g^{-1} \rm cm^{-2}$. However, the actual value
of $\kappa_{850}$ is rather uncertain. \citet{James2002} used SCUBA
observations to conclude to a value $\kappa_{850} = 0.7\pm0.2\ \rm
g^{-1} \rm cm^{2}$ but their data show a spread in values for
individual galaxies by a factor of two. Worse, \citet{Alton2000} found
a three times higher value for the galaxy NGC~891, whereas
\citet{Agladze1996} in the laboratory came up with a value about two
times higher than that of \citet{James2002}.

We can make an estimate of $\kappa_{850}$ in M~51. When we increase
the fraction of graphite in the grains, a higher value for the dust
emission coefficient is obtained. With a higher dust emission
coefficient, the surface mass density will be lower. For $\kappa_{850}
= 1.2\ \rm g^{-1} \rm cm^{2}$, the peak surface densities as shown in
Fig.~\ref{surfdens} will be 0.65 (no CO contribution) and 0.56 (12
percent CO). In table \ref{dustgasratios}, gas-to-dust ratios are
shown for four different gas mixtures and for a luminosity of
$L=1.1\cdot 10^{11}\rm L_\odot$.  A canonical gas-to-dust ratio of
100$\pm$26 is obtained with a dust emission coefficient
$\kappa_{850}=1.2\ \rm g^{-1} \rm cm^{2}$, somewhat less than twice
the value found by \citet{James2002}.

\begin{figure}[]
\centerline{\includegraphics[height=65mm,clip=]{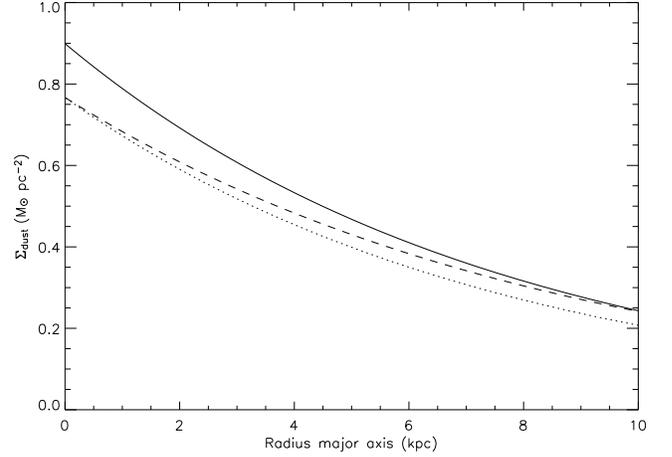}}
\caption{The dust surface density profiles for a graphite and silicate
ratio of 0.5:0.5 and $kappa_{850} = 1.2\ \rm g^{-1} \rm cm^{2}$; no CO
contribution (solid line), 12 percent CO contribution (dotted line)
and decreasing CO contribution (dashed line). }
\label{surfdens}
\end{figure}

\begin{table}
\caption[]{Gas to dust-ratios}
\begin{flushleft} 
\begin{tabular}{llll} 
\hline
\noalign{\smallskip}
silicate:graphite  & $\kappa_{850}\ (\rm g^{-1} \rm cm^{2})$  &  dust:gas \\
\noalign{\smallskip}
\hline
\noalign{\smallskip}
0.5:0.5  & 0.9  & 73$\pm$17  \\
0.4:0.6  & 1.0  & 82$\pm$21  \\                 
0.3:0.7  & 1.1  & 91$\pm$24  \\
0.2:0.8  & 1.2  & 100$\pm$26 \\
\noalign{\smallskip}
\hline
\end{tabular}
\end{flushleft}
\label{dustgasratios}
\end{table}

\section{Summary}

\begin{enumerate}
\item{SCUBA 850~$\mu$m observations of M~51 show a well-delineated
spiral structure superimposed on a prominent extended exponential base
disk.}
\item{The observed base disk is evidence of an underlying exponential
density distribution of cold dust in M~51. This reinforces the
suggestion that the cold dust in spiral galaxies traces the total
hydrogen density, i.e. the sum of H$_2$ and HI}
\item{While throughout M~51 a radial temperature gradient occurs,
because the stars are more centrally concentrated than the dust, this
gradient by itself can not be the origin of the exponential nature of
the observed 850~$\mu$m disk.}  
\item{A reasonable estimate of the dust emission coefficient, based on
a canonical gas to dust ratio of $100\pm26\ $, is $\kappa_{850} = 1.2\
\rm g^{-1} cm^{2}$.}
\item{We find a stellar scale-length $h_s$ of 3.15 kpc and a dust
scale-length $h_d$ that ranges from 7.65 to 8.65 kpc.}
\item{We find a typical central face-on optical depth
$\tau_{V,f}=4.0$.}
\end{enumerate}

\begin{acknowledgements} 
This research has made use of the NASA/IPAC Extragalactic Database
(NED) which is operated by the Jet Propulsion Laboratory, California
Institute of Technology, under contract with the National Aeronautics
and Space Administration. We want to thank Eric Bell for a helpful
discussion about dust opacities. RT thanks Ron Allen and Leo Blitz for
an invigorating discussion on the nature of the exponential dust disk.
We also thank S.\ Bianchi and the referee, R.J.\ Tuffs, for their
insightful comments and helpful suggestions which have improved this
paper.
\end{acknowledgements}

\end{document}